\documentstyle[twoside,fleqn,espcrc2]{article}

\newcommand{\be}{\begin{equation}}
\newcommand{\ee}{\end{equation}}
\newcommand{\bea}{\begin{eqnarray}}
\newcommand{\eea}{\end{eqnarray}}
\newcommand{\nn}{\nonumber \\}
\newcommand{\p}[1]{(\ref{#1})}

\font\mybb=msbm10 at 10pt
\def\bb#1{\hbox{\mybb#1}}
\def\R {\bb{R}}

\newcommand{\AmS}{{\protect\the\textfont2
  A\kern-.1667em\lower.5ex\hbox{M}\kern-.125emS}}

\title{M-branes at angles}

\author{P.K. Townsend\address{DAMTP, University of Cambridge,\\ 
       Silver St., Cambridge, U.K.}}
       
\begin{document}

\begin{abstract}
Supersymmetric configurations of non-orthogonally intersecting M-5-branes can be
obtained by rotation of one of a pair of parallel M-5-branes. Examples
preserving 1/4, 3/16 and 1/8 supersymmetry are reviewed.
\end{abstract}

\maketitle

\section{INTRODUCTION}

In contrast to the current rapid evolution in our understanding of the
microscopic degrees of freedom of M-theory, the macroscopic picture has changed
very little over the past year. The `basic' ingredients remain D=11
supergravity and its 1/2-supersymmetic solutions: the M-wave, the
M-2-brane, the M-5-brane and, in the $S^1$ Kaluza-Klein
vacuum, the M-monopole (i.e. Euclidean-Taub-Nut times 7-dimensional
Minkowski). The novelty has been in the way that these ingredients have
been combined to form new `intersecting-brane' configurations, preserving
some smaller fraction of supersymmetry, and in the remarkable insights into
quantum field theory that these configurations have led to. I shall have
nothing to say here about these applications of intersecting brane
configurations. Instead, I shall discuss one aspect of attempts towards a
classification of them: the
conditions for partial preservation of supersymmetry by configurations of {\sl
non-orthogonally} intersecting M-branes. 

While it is not
difficult to see why orthogonal intersections of branes can partially preserve
supersymmetry, it is less obvious that this is also true for non-orthogonal
intersections. This possibility was first pointed out by Berkooz, Douglas and
Leigh \cite{BDL}, who also provided several D-brane examples, which they
interpreted as rotations within an SU(n) subgroup of SO(2n). Further examples,
interpretable as  rotations within an $Sp(2)$ subgroup of $SO(8)$, were given by
Gauntlett, Gibbons, Papadopoulos and the author \cite{GGPT} and shown to be
related via M-theory dualities to 8-dimensional hyper-K{\"a}hler manifolds. I
shall  review both cases here although the interpretation as rotations within
subgroups of reduced holonomy and the connection with (hyper)complex geometry
will not be explained. Instead I shall concentrate on clarifying some aspects of
the Dirac matrix algebra needed to determine the fraction of supersymmetry
preserved by rotated M-brane configurations, i.e. `M-branes at angles'. 

It is convenient to start with two parallel M-5-branes. Essentially no 
generality is lost by considering only M-5-branes, and any non-parallel
configuration of two intersecting M-5-branes can obtained by some rotation of one
of them from the parallel configuration. We may assume that the two parallel
M-5-branes lie in the 12349 5-plane. This configuration can be represented by 
the array
$$
\begin{array}{lcccccccccc}
M: & 1 & 2 & 3 & 4 & - & - & - & - & 9 & - \\
M: & 1 & 2 & 3 & 4 & - & - & - & - & 9 & - 
\end{array}
$$
and is associated with the constraint
\be
\label{1}
\Gamma_{091234}\epsilon = \epsilon
\ee
where $\epsilon$ is the asymptotic value of a Killing spinor of the
corresponding supergravity solution. Here we shall refer to the solutions of 
this and similar algebraic equations as `Killing spinors'. Since
$\Gamma_{091234}$ is traceless and squares to the identity, the space of
solutions of \p{1} is 16-dimensional, i.e. the configuration preserves 1/2
supersymmetry. We now wish to rotate the second M-5-brane away from the 12349
5-plane to give a configuration of two intersecting M-5-branes. If the spinor
representation of the rotation matrix is $R$ then the constraint imposed by the
presence of the second M-5-brane is \cite{BDL} 
\be
\label{2}
R\Gamma_{091234}R^{-1}\epsilon = \epsilon\, .
\ee

Although $R$ may be a general $SO(10)$ matrix, the relative orientation of the
second M-5-brane is actually determined by only five angles. To see this we
observe \cite{GGPT} that the orientation of each M-5-brane is determined by a set
of five linearly independent normals, so that the relative orientation is
determined by the $5\times 5$ matrix $M$ of inner products of one set of normals
with the other. Each set can be taken to be orthonormal. The remaining freedom 
in the choice of the two sets of normals allows the diagonalization of
$M$ by the action of $O(5;\R)\times O(5;\R)$. The diagonal entries are the
cosines of the five angles. In the spinor represention, the rotation
matrix can be chosen to be\footnote{We use a base 11 arithmetic in which
$\natural$ is the symbol for the number 10. Since the standard notation for
integers is based on the number of angles (none for $0$, one for $1$, etc) a
symbol with ten angles would be ideal although impractical. The
symbol `$\natural$', to be pronounced `ten', seems to be a reasonable 
compromise. In any case it is certainly a `natural' choice.}
\be
\label{3}
R = e^{{1\over2}[\vartheta \Gamma_{15} + \psi\Gamma_{26} + 
\varphi\Gamma_{37} + \rho\Gamma_{48} + \zeta\Gamma_{9\natural}]}
\ee 
where $\vartheta,\psi,\varphi,\rho$ and $\zeta$ are the five angles. Note that
$\Gamma_{091234}R^{-1} = R\Gamma_{091234}$, so that \p{2} becomes
$R^2\Gamma_{091234}\epsilon=\epsilon$. In view of \p{1}, this means that the
constraint imposed on Killing spinors by the presence of the second M-5-brane is
equivalent to
\be
\label{4}
[R^2 -1]\epsilon =0
\ee
with $R$ given by \p{3}. 

What we now have to do now is to determine for a given configuration, specified
by the five angles, the number of solutions to the simultaneous equations \p{1}
and \p{4}, and hence the fraction $\nu$ of supersymmetry preserved by the
configuration. Ultimately, we would like to determine $\nu$ as a (discontinuous)
function of $\vartheta,\psi,\varphi,\rho,\zeta$, but this will not be attempted
here. Instead, some particular cases preserving 1/4, 3/16 and 1/8 supersymmetry
will be reviewed. 

\section{1/4 SUPERSYMMETRY}

We begin the analysis by considering the case of a rotation by an angle
$\vartheta$ in the 15 plane, for which
\be
\label{5}
R^2 -1= [e^{\vartheta\Gamma_{15}} -1]\, .
\ee
This has a zero eigenvalue only for $\vartheta=0$ (mod $2\pi$). Thus, a rotation
in the 15-plane breaks supersymmetry. We therefore move on to consider a
simultaneous rotation in the 15 and 26 planes. In this case one easily sees that $(R^2-1)$ can have zero eigenvalues only if
$\psi=\pm\vartheta$. The sign is irrelevant, so set $\psi = \vartheta$ (and
$\varphi=\rho=\zeta=0$) in \p{3} to get the constraint
\bea
0&=& \big[e^{\vartheta\,[\Gamma_{15} + \Gamma_{26}]} -1\big]\epsilon\nn
 &=& \sin\theta\, \Gamma_{15} e^{\vartheta \Gamma_{15}} (1-
\Gamma_{1526})\epsilon\, .
\label{6}
\eea
This is an identity if $\vartheta=0$, as expected, but it is also an 
identity if 
$\vartheta=\pi$. The reason for this is that while a rotation by $\pi$ in the
15-plane just converts the M-5-brane into an anti-brane, the simultaneous
rotation by $\pi$ in the 26-plane returns it to the original M-5-brane. 
If $\sin\vartheta$ is non-zero then \p{6} is satisfied only if
\bea
\label{7 }
\epsilon &=& \Gamma_{1526}\epsilon \equiv
 \Gamma_{034569} \Gamma_{012349}\epsilon \nn
         &=& \Gamma_{034569}\epsilon
\eea
where the last line follows from \p{1}. But this is just the constraint
associated with an M-5-brane in the 34569 5-plane. Thus, as shown
originally in the context of D-branes \cite{BDL}, whatever the value of the angle $\theta$ (other than
zero mod $2\pi$) the rotated configuration preserves the same fraction of
supersymmetry as the orthogonal intersection of two M-5-branes on a 3-plane,
represented by the array
$$
\begin{array}{lcccccccccc}
M: & 1 & 2 & 3 & 4 & - & - & - & - & 9 & - \\
M: & - & - & 3 & 4 & 5 & 6 & - & - & 9 & - 
\end{array}
$$
This fraction is 1/4 \cite{paptown,AT}. In fact, the orthogonal intersection of
any pair of branes either breaks all supersymmetry or preserves 1/4
supersymmetry. One way to understand why certain rotations away from
orthogonality can preserve the 1/4 supersymmetry of orthogonal intersections is
that the non-orthogonal rotations are related to the orthogonal ones by duality
\cite{cvetic}. This explanation is, of course, not available for rotations which
preserve less than 1/4 supersymmetry, although orthogonal intersections of more
than two branes can yield 1/8, 1/16, and even \cite{berg} 1/32 supersymmetry.

The next case to consider in a systematic analysis of rotations of two
M-5-branes would be simultaneous rotations in three independent planes. It is
straightforward to see that partial preservation of supersymmetry in
this case requires the three angles to sum to zero (mod $2\pi$ and for an
appropriate choice of the signs). An example of this type was given in
\cite{BDL} in which two IIA 6-branes intersect on a 3-plane: the intersection
preserves 1/8 supersymmetry if the relative orientation is obtained by a
rotation within an
$SU(3)$ subgroup of
$SO(6)$. The M-5-brane dual of this configuration is two M-5-branes intersecting
non-orthogonally on a 2-plane. The orthogonal intersection of this type would
break all supersymmetry, but it cannot be reached by a rotation in which the
three angles sum to zero. We shall skip the detailed analysis of this case and
move on to simultaneous rotations in four independent planes. A complete
analysis of the latter case will not be attempted either. Instead, some examples
of special interest will be described, starting with one that preserves 3/16
supersymmetry, a fraction that cannot be realized by orthogonal intersections.

\section{3/16 SUPERSYMMETRY}

By means of an explicit calculation in a particular representation of the Dirac 
matrices, the rotation matrix
\be
\label{8}
R = e^{{1\over2}\vartheta [\Gamma_{15} + \Gamma_{26} + 
\Gamma_{37} + \Gamma_{48}]}
\ee 
was shown in \cite{GGPT} to lead to preservation of 3/16 supersymmetry. Here, I
shall give an alternative, representation-independent, proof of the 3/16
supersymmetry. We first note that
\be
\label{9}
R^2 = \prod_{i=1}^4 \big(\cos\vartheta + \sin\vartheta \Gamma_{i\, i+4}\big)\, .
\ee
Carrying out the multiplications, we find that
\bea
\label{10}
R^2 -1 = \sin^4\vartheta (\Gamma_{12345678} -1) + {1\over8}\Sigma^2 \sin^2
2\vartheta && \nn
+ {1\over2}\sin 2\vartheta (\cos^2\vartheta
-\Gamma_{12345678} \sin^2\vartheta)\  && 
\eea
where
\be
\label{11}
\Sigma = [\Gamma_{15} + \Gamma_{26} + \Gamma_{37} + \Gamma_{48}]\, .
\ee
As expected, $R^2=1$ when $\vartheta=0,\pi$. When $\vartheta=\pi/2$ we have
$R^2 = \Gamma_{12345678}$, so that $\epsilon$ must
satisfy $\Gamma_{12345678}\epsilon=\epsilon$. In view of \p{1} this is 
equivalent to the constraint
\be
\label{14}
\Gamma_{056789}\epsilon =\epsilon\, .
\ee
This corresponds to the orthogonal intersection of two M-5-branes on a line,
summarized by the array
$$
\begin{array}{lcccccccccc}
M: & 1 & 2 & 3 & 4 & - & - & - & - & 9 & - \\
M: & - & - & - & - & 5 & 6 & 7 & 8 & 9 & - 
\end{array}
$$
This is as expected because a $\vartheta=\pi/2$ rotation yields precisely
this configuration, which is known to preserve 1/4 supersymmetry \cite{GKT}.

For other values of $\vartheta$ we proceed as follows. The matrix $\Sigma$
satisfies
\be
\label{15}
\Sigma^2 + 4 = 2\Lambda
\ee
where
\bea
\label{16}
\Lambda &=& \Gamma_{1526} + \Gamma_{1537} + \Gamma_{1548} \nn
 && +\, \Gamma_{2637} + \Gamma_{2648} + \Gamma_{3748}\, .
\eea
The matrix $\Lambda$ satisfies
\be
\label{17}
\Lambda^2 + 4\Lambda -6 = 6\Gamma_{12345678}\, .
\ee
It follows that $\Lambda$ and $\Gamma_{12345678}$ are simultaneously
diagonalizable. Moreover
\be
\label{18}
\Gamma_{12345678}\Lambda \equiv \Lambda\, ,
\ee
so all eigenvalues of $\Lambda$ in the ($-$) eigenspace of $\Gamma_{12345678}$
vanish. In the ($+$) eigenspace \p{17} yields eigenvalues $-6$ and $2$. 
Since $\Lambda$ is traceless the relative multiplicity of these non-zero
eigenvalues is $1:3$.

It is convenient to now introduce the constraint \p{1}. As $\Gamma_{091234}$
commutes with $\Lambda$ (and with $\Gamma_{12345678}$) we can write $\Lambda
=\Lambda_+ +\Lambda_-$, where $\Lambda_\pm$ are the projections of $\Lambda$ 
onto the ($\pm$) eigenspaces of $\Gamma_{091234}$. Now, both $\Lambda$ and
$\Gamma_{091234}\Lambda$ are traceless, so $\Lambda_+$ and $\Lambda_-$ are
traceless too. The relative multiplicities of the eigenvalues of $\Lambda_+$ are
therefore the same as those of $\Lambda$. In other words, the eigenspinors of
$\Lambda$ in the 16-dimensional eigenspace of $\Gamma_{091234}$ with eigenvalue
$+1$ are again $0,-6,2$ and the multiplicities are $8,2,6$. This translates to
the following eigenvalues of $\Sigma^2$:
\be
\label{19}
 -4\, (8,-) \, ,\qquad -16\, (2,+)\, ,\qquad 0\, (6,+)\, .
\ee
The numbers in parentheses are the multiplicities of eigenvalues for
eigenspinors satisfying \p{1} and the sign is the sign of the eigenvalue of
$\Gamma_{12345678}$. Now \p{10} can be rewritten as
\bea
R^2_+ -1 &=& {1\over2}\Sigma\sin 2\vartheta (\cos 2\vartheta +
{1\over4}\Sigma
\sin 2\vartheta)\ \label{20}\\
R^2_- -1 &=& -2\sin\vartheta \bigg[\sin\vartheta 
-{1\over2}\Sigma\cos\vartheta \nn 
&&-{1\over8}(\Sigma^2 +4) \cos\vartheta\sin 2\vartheta\bigg] 
\label{21}
\eea
where the $(\pm)$ subscript indicates the eigenspace of $\Gamma_{12345678}$.
The zero eigenvalues of $\Sigma^2$ clearly correspond to zero eigenvalues of 
$R^2 -1$ and it is straightforward to check that the non-zero eigenvalues of $\Sigma^2$ do not. There are therefore precisely 6 independent simultaneous solutions of \p{1} and \p{4}, implying preservation of 3/16 supersymmetry.

\section{1/8 SUPERSYMMETRY}
 
We shall conclude this analysis by considering a case that, in general,
preserves only 1/8 supersymmetry but which includes both the previous
$\nu=3/16$ and $\nu=1/4$ cases as special limits. 
We start from the rotation matrix
\be
\label{22}
R= e^{{1\over2}\vartheta(\Gamma_{15} + \Gamma_{26}) + {1\over2}
\psi(\Gamma_{37} + \Gamma_{48})}\, .
\ee
We compute
\bea
R^2 &&= [\cos^2\vartheta + {1\over2}\mu\sin 2\vartheta + A
\sin^2\vartheta]\times \nn
&& [\cos^2\psi + {1\over2}\nu\sin 2\psi + B \sin^2\psi]
\label{24}
\eea
where
\be
\label{25}
\mu = (\Gamma_{15}+ \Gamma_{26})\, , \qquad \nu=(\Gamma_{37}+\Gamma_{48})\, ,
\ee
and
\be
\label{26}
A = {1\over2} (\mu^2 +2)\, , \qquad B = {1\over2} (\nu^2 +2)\, .
\ee
It is straightforward to show that $A$ and $B$ are both traceless and square to
the identity. Moreover, $A$ and $B$ commute with $\Gamma_{12345678}$ and are such
that both $\Gamma_{12345678}A$ and $\Gamma_{12345678}B$ are traceless, so the
projections of $A$ and $B$ onto the eigenspaces of $\Gamma_{12345678}$ are traceless too. Finally, we observe that
\be
\label{27}
AB =\Gamma_{12345678}\, ,
\ee
so that the product of the eigenvalue $a$ of $A$ with $b$ of $B$ is $\pm1$,
according to the eigenvalue of $\Gamma_{12345678}$. 

With this information we compute that for $ab=1$
\be\label{28}
$$
R^2 = \left\{
\begin{array}{ll} 
1& a=1 \nn
e^{\pm 2i(\vartheta +\psi)}\ {\rm or}\ 
  e^{\pm 2i(\vartheta -\psi)} 
& a=-1 
\end{array} \right.                
$$
\ee
where the multiplicities of the $a=\pm 1$ eigenvalues of $A$ are both equal to 8.
For
$ab=-1$,
\be\label{29}
$$
R^2 = \left\{
\begin{array}{ll} 
e^{\pm 2i\vartheta}& a=1 \nn
e^{\pm 2i\psi}& a=-1 
\end{array} \right.               
$$
\ee
where the multiplicities of $a=\pm 1$ eigenvalues of $A$ are again both equal to
8.

At this point we may introduce the constraint \p{1}. We observe that the
traceless matrices $A$ and $B$ commute with $\Gamma_{091234}$ and are such that
both $\Gamma_{091234}A$ and $\Gamma_{091234}B$ are traceless too. The same
reasoning as before then shows that the projections of $A$ and $B$
onto the subspace satisfying \p{1} have the same eigenvalues and multiplicities
as $A$ and $B$ themselves. This effectively reduces the multiplicities of 
of the $a=+1$ and $a=-1$ eigenvalues of $A$ within each $\Gamma_{12345678}$
eigenspaces from 8 to 4. It follows immediately that for generic angles
$\vartheta$ and $\psi$ the constraints $\Gamma_{091234}\epsilon =\epsilon$ and
$(R^2-1)\epsilon=0$ have just 4 common solutions, corresponding to $a=1$ in
\p{28}. Thus, the generic rotation with $R$ of the form \p{22} leads to 4/32,
or 1/8, supersymmetry. 

For special values of $\vartheta$ and $\psi$ there can be additional solutions
of $(R^2-1)\epsilon=0$. For example, when either $\vartheta$ or $\psi$ vanishes
$(R^2-1)$ has four more zero eigenvalues in the subspace satisfying \p{1}.
These are the 1/4 supersymmetric solutions discussed above in which two 
M-5-branes intersect on a 3-plane. When neither $\vartheta$ nor $\psi$ are zero
but either $\vartheta +\psi$ or $\vartheta-\psi$ is (mod $2\pi$), then four of
the eight eigenvalues of $(R^2-1)$ corresponding to the $a=-1$ eigenvalue of $A$
in \p{28} vanish. It is also true, as we saw above, that only four of these
eight eigenvalues of $(R^2-1)$ correspond to eigenspinors satisfying \p{1}.
Unfortunately, it is not obvious from this analysis how many of the latter  are
zero modes of $(R^2-1)$. In fact, just two of them will be because for the given
choice of $\vartheta$ and $\psi$ the matrix $R$ reduces (up to irrelevant choices of signs of the two angles) to \p{8}, which we earlier showed to lead  to 3/16
supersymmetry. Note that if were not for this ambiguity in the current
analysis the previous analysis of the 3/16 superymmetric case would have been
unnecessary. The ambiguity does not arise for the special case in which either
$\vartheta=\psi=\pm \pi/2$ or $\vartheta=-\psi =\pm \pi/2$ since in that case
all eight eigenvalues of $(R^2-1)$ corresponding to $a=-1$ in \p{28} vanish and
the issue of which four to choose does not arise. Whichever four we take to
satisfy \p{1} we have a total of 8 common solutions to the M-5-brane constraints
and hence 1/4 supersymmetry. This case is of course just the {\sl orthogonal}
intersection on a line of two M-5-branes.

\section{CONCLUSIONS}

I have shown in detail how configurations of non-orthogonally intersecting
M-5-branes may preserve 1/4, 3/16, or 1/8 supersymmetry. These results were
obtained originally in \cite{BDL} and \cite{GGPT} by slightly different
methods. It is likely that there is a further supersymmetric configuration
obtained by a simultaneous rotation in four independent planes, corresponding to
rotation within a $Spin(7)$ subgroup of $SO(8)$. If so it should preserve 1/16
supersymmetry. It may also be possible to realize this fraction by a
simultaneous rotation in five independent planes, corresponding to an
$SU(5)$ rotation in $SO(10)$. It would be desirable to have a comprehensive
analysis along the above lines to verify this. At any rate, this would appear to
be a necessary step for a classification of all supersymmetric 
M-theory configurations. 
\vskip 0.5cm 

\noindent
{\bf Acknowledgements}: I am grateful to Eric Bergshoeff, Michael Douglas,
Jerome Gauntlett, Gary Gibbons and George Papadopoulos for many discussions on
intersecting branes.

\end{document}